\documentclass{aa}
\usepackage{graphicx,amssymb,subfigure}
\usepackage[normalem]{ulem}
\usepackage{txfonts}
\begin{document}

\newcommand{\wav}[1]{$\lambda#1\,{\rm cm}$}  
\newcommand{\wwav}[2]{$\lambda\lambda#1,#2\,{\rm cm}$}  
\newcommand{\wwwav}[3]{$\lambda\lambda#1,#2,#3\,{\rm cm}$}  
\newcommand{\Bt}{\,B_\mathrm{tot}}  
\newcommand{\Btpa}{\,B_\mathrm{tot\parallel}}
\newcommand{\Btpe}{\,B_\mathrm{tot\perp}}
\newcommand{\Br}{\,B_\mathrm{reg}}  
\newcommand{\Brpa}{\,B_{\parallel}}
\newcommand{\Brpe}{\,B_{\perp}}
\newcommand{\bt}{\,b} 
\newcommand{\btpa}{\,b_{\parallel}}
\newcommand{\btpe}{\,b_{\perp}}

\newcommand{\cs}{c_{\rm s}}
\newcommand{\deriv}[2]{\displaystyle\frac{\partial #1}{\partial #2} }
\newcommand{\HI}{{\rm H\,\scriptstyle I}}
\newcommand{\N}{{N}}
\newcommand{\n}{{n}}
\newcommand{\ncr}{n_{\rm cr}}
\newcommand{\nel}{n_{\rm e}}
\newcommand{\ntot}{n_{\rm t}}
\newcommand{\Pgas}{{{\cal P}_{\rm gas}}}
\newcommand{\RM}{{\rm RM}}
\newcommand\sfrac[2]{{\textstyle{\frac{#1}{#2}}}}

%
%
\newcommand{\cm}{\,{\rm cm}}
\newcommand{\cmcube}{\,{\rm cm^{-3}}}
\newcommand{\dyn}{\,{\rm dyn}}
\newcommand{\erg}{\,{\rm erg}}
\newcommand{\Jy}{\,{\rm Jy}}
\newcommand{\Jyb}{\,{\rm Jy/beam}}
\newcommand{\kms}{\,{\rm km\,s^{-1}}}
\newcommand{\mJy}{\,{\rm mJy}}
\newcommand{\mJyb}{\,{\rm mJy/beam}}
\newcommand{\K}{\,{\rm K}}
\newcommand{\kpc}{\,{\rm kpc}}
\newcommand{\Mpc}{\,{\rm Mpc}}
\newcommand{\mG}{\,{\rm mG}}
\newcommand{\mkG}{\,\mu{\rm G}}
\newcommand{\MHz}{\, {\rm MHz}}
\newcommand{\Msol}{\,{\rm M_\odot}}
\newcommand{\p}{\,{\rm pc}}
\newcommand{\radm}{\,{\rm rad\,m^{-2}}}
\newcommand{\s}{\,{\rm s}}
\newcommand{\yr}{\,{\rm yr}}

\title{\bf {Magnetic fields in barred galaxies III:}\\
The southern peculiar galaxy NGC 2442}

\author{
Julienne Harnett \inst{1,3}
\and
Matthias Ehle \inst{2}
\and
Andrew Fletcher \inst{3}
\and
Rainer Beck \inst{3}
\and
Raymond Haynes \inst{4}
\and
Stuart Ryder \inst{5}
\and
Michael Thierbach \inst{3}
\and
Richard Wielebinski \inst{3}
}

\offprints{J.~Harnett,\email{jules@eng.uts.edu.au}}
\authorrunning{J.I. Harnett et al.}
\titlerunning{Barred spirals: NGC 2442}

\institute{ University of Technology Sydney, PO Box 123, Broadway
  2007, NSW, Australia
  \and XMM-Newton Science Operations Centre,
  European Space Agency, Villafranca, PO Box 50727, 28080 Madrid,
  Spain
  \and Max-Planck-Institut f\"ur Radioastronomie, Auf dem
  H\"ugel 69, 53121 Bonn, Germany
  \and University of Tasmania, GPO Box
  252-21, Hobart 7001, Tasmania, Australia
  \and Anglo-Australian
  Observatory, P.O.Box 296, Epping, NSW 1710, Australia }

\date{Received ...; accepted ...}

\abstract{ Observations of the southern peculiar galaxy \object{NGC 2442} with
  the Australia Telescope Compact Array in total and linearly
  polarized radio continuum at $\lambda$6~cm are presented and
  compared with previously obtained H$\alpha$ data. The distribution
  of polarized emission, a signature of regular magnetic fields,
  reveals some physical phenomena which are unusual among spiral
  galaxies. We find evidence for tidal interaction and/or ram pressure
  from the intergalactic medium compressing the magnetic field at the
  northern and western edges of the galaxy. The radial component of
  the regular magnetic field in the northern arm is directed away from
  the centre of the galaxy, a finding which is in contrast to the
  majority of galaxies studied to date. The oval distortion caused by
  the interaction generates a sudden jump of the magnetic field
  pattern upstream of the inner northern spiral arm, similar to
  galaxies with long bars. An unusual ``island'' of strong regular
  magnetic field east of the galaxy is probably the brightest part of
  a magnetic arm similar to those seen in some normal spiral galaxies,
  which appear to be phase-shifted images of the preceding optical
  arm.  The strong magnetic field of the ``island'' may indicate a
  past phase of active star formation when the preceding optical arm
  was exposed to ram pressure.   }

\maketitle

\section{Introduction}

More than twenty years of observations and modelling have shown that
large-scale, or regular, magnetic fields pervade the interstellar
medium in all spiral galaxies (Beck~\cite{Beck00}, \cite{Beck02}).
Observing linearly polarized radio emission is a well-established
method for investigating the strength and structure of such regular
magnetic fields, which are often controlled by the dynamics of the
interstellar gas.

Large-scale magnetic fields in spiral galaxies lie predominantly in
the plane of the discs.  In normal (i.e. ``non-barred'') spirals, the
morphology is usually qualitatively similar to the optical appearance
of the host galaxy except that the strongest regular fields often lie
in the inter-arm regions (Beck \& Hoernes~\cite{Beck96}). These
``magnetic arms'' may be caused by enhanced magnetic diffusion in the
optical arms (Moss \cite{Moss98}) and/or by enhanced dynamo action in
the inter-arm regions (Rohde \& Elstner \cite{Rohde98}, Shukurov
\cite{Shukurov98}).

The presence of a stellar bar in a spiral galaxy, on the other hand,
results in highly non-circular motions of the gas and stars in a
galaxy. Strong deflection of gas streamlines along shock fronts in the
bar region and significant compression of the gas have been predicted
by Athanassoula (\cite{Athanassoula92}), Piner et al.
(\cite{Piner95}), Lindblad et al.  (\cite{Lindblad96}) and Englmaier
\& Gerhard (\cite{Englmaier97}).  Gas in the bar region rotates faster
than the bar pattern itself and compression regions develop, traced by
dust lanes. This gas inflow is hard to observe spectroscopically. In
addition, the processes involved are likely to be interdependent as
collisions of dense gas clouds and shocks in the bar potential may be
modified by strong, regular magnetic fields.  Theoretical models have
recently begun to address the relationship between magnetic fields and
gas flows in barred spirals (Moss et al.~\cite{Moss01}).

Radio observations of the SBc galaxy NGC~1097 indicate that the
regular magnetic fields in barred galaxies differ markedly
from those we see in non-barred galaxies (Beck et al.~\cite{Beck99}).
Instead of exhibiting an open spiral pattern that is qualitatively
similar to the optical morphology, the regular field appears to be
tightly bound to the local gas flow in the bar. Gas inflow along the
compression region may fuel star formation in a dense, inner,
circum-nuclear ring, which is also delineated by enhancement of the
total, mostly turbulent magnetic field. In the circum-nuclear ring
regular magnetic fields appear spiral in shape and so they probably
decouple from the gas flow. As a result, the regular magnetic fields
may facilitate angular momentum transfer which results in funnelling of
the circum-nuclear gas toward the active nucleus.

We have undertaken radio continuum observations of 20 barred galaxies
with the Effelsberg 100m telescope, the Very Large Array (VLA)
operated by the NRAO \footnote{The National Radio Astronomy
Observatory is a facility of the National Science Foundation operated
under cooperative agreement by Associated Universities, Inc.}  and the
Australia Telescope Compact Array (ATCA). \footnote{The Australia
Telescope Compact Array is part of the Australia Telescope which is
funded by the Commonwealth of Australia for operation as a National
Facility managed by CSIRO.} As this project consisted of the first
systematic search for magnetic fields in barred spiral galaxies, we
chose a sample of galaxies with prominent optical bars. Ten of the
galaxies were observed with the VLA and Effelsberg telescopes, the
rest with the ATCA telescope. We are continuing our research into the
magnetic fields and dynamic processes in the circum-nuclear regions of
barred galaxies with detailed observations of a sub-sample of the
original galaxies. Details are given by Beck et al.  (\cite{Beck02}).

In this paper, we discuss the southern peculiar galaxy NGC\,2442.

\section{NGC\,2442}

Basic information about NGC\,2442 is given in Table~\ref{tab:gal}.
Following Ryder et al. (\cite{Ryder01}), we use a distance of
15.5~Mpc, at which $1\arcsec$ corresponds to 75~pc along the major
axis and 82~pc along the minor axis.

The striking appearance of NGC\,2442 (see Fig.~\ref{fig:i6}) is
predominantly due to the deformed outer spiral arms so that the inner
spiral arms look like a huge bar in the NE-SW direction. The true bar,
leading to the galaxy's optical classification (de Vaucouleurs et
al.~\cite{deVaucouleurs91}), is a small feature running EW, $\sim
66\arcsec$ (5.0~kpc) long, centred on the nucleus. According to Bajaja
et al.  (\cite{Bajaja95}, \cite{Bajaja99}) and Mihos \& Bothun
(\cite{Mihos97}) there is also a weak, elliptical circum-nuclear ring
of molecular gas and star formation that has the same major axis
orientation as the bar and a radius of $\sim 12\farcs 5$.

Two spiral arms emerge from the ends of the bar and proceed
symmetrically along the NE-SW direction for $\sim 2\arcmin$, but then
become completely asymmetric.  The northern arm is well-developed, but
elongated and bisected by a prominent dust lane. It is bent back on
itself by over $90\degr$. The southern arm is less conspicuous,
broader and traversed by many apparently chaotic dust lanes (see
Fig.~\ref{fig:i6}), and bends by about $180\degr$ towards the
northeast, reaching the left edge of Fig.~\ref{fig:i6}.

The distorted morphology hints at recent tidal interaction or ram
pressure stripping as the galaxy traverses the intra-cluster medium
(Mihos \& Bothun~\cite{Mihos97}; Ryder et al.~\cite{Ryder01}).
Imprints of non-circular rotation are evident from the velocity field,
especially around the northern spiral arm (Bajaja et al.
\cite{Bajaja99}, Houghton~\cite{Houghton98}).

\begin{table}[ht]
  \begin{center}
    \caption{Properties of NGC\,2442}
    \label{tab:gal}
    \begin{tabular}{ll}
      \hline\hline
      \noalign{\smallskip}
      Parameter \\
      \noalign{\smallskip}
      \hline
      \noalign{\smallskip}
      Right ascension (J2000) & $\rm 07^h 36^m 23\fs 9$\\
      Declination (J2000) & $-69\degr 31\arcmin 50\arcsec$ \\
      Type & SBbc(rs) II\\
      Major diameter $D_{25}$ (\arcmin ) & 5.5\\
      Distance (Mpc) & 15.5$ ^{1}$ \\
      Inclination ($0\degr$ is face on) & $24\degr$ $^{2,3}$ \\
      PA of major axis & $40\degr$ $^2$ \\
      \noalign{\smallskip} \hline
      \noalign{\smallskip}
    \end{tabular}\\
  \end{center}
	$^{1}$ Ryder et al. (\cite{Ryder01}). \\
  $^{2}$ Bajaja et al. (\cite{Bajaja99}). However, Ryder (\cite{Ryder95})
	argues that the galaxy is more highly inclined.\\
  All other data are from de Vaucouleurs et al. (\cite{deVaucouleurs91})\\
\end{table}

Recent work referred to by Ryder et al.  (\cite{Ryder01}) has shown
that the group of galaxies in Volans associated with NGC\,2442
contains more than the three other galaxies originally named by Garcia
(\cite{Garcia93}): NGC\,2397, NGC\,2434, and PGC\,20690. The
perpetrator of the alleged interaction, however, remains unclear,
particularly in view of the discovery of a large gas cloud with
obvious streaming structures, HIPASS\,J0731~-~69, apparently
associated with NGC\,2442 and extended to the north and west (Ryder et
al.  \cite{Ryder01}).

Earlier observations in \ion{H}{i} (Bajaja \& Martin~\cite{Bajaja85})
provided mass and systemic velocity estimates, while
\element[][12]{CO}(1--0) observations with SEST (Harnett et
al.~\cite{Harnett91}; Bajaja et al.~\cite{Bajaja95}) provided the
velocity field of the molecular gas.

The \element[][12]{CO}(1--0) distribution correlates remarkably well
with the Molonglo Observatory Synthesis telescope (MOST)
$\lambda$35~cm emission presented by Harnett (\cite{Harnett84}). Both
SEST and the MOST have angular resolutions $\sim 43\arcsec$.  ATCA
images in \ion{H}{i} and $\lambda$20~cm continuum
(Houghton~\cite{Houghton98}) also posses similar morphological
characteristics except that \ion{H}{i} is weak in the galaxy's centre.

Between 5~GHz and 408~MHz, the radio spectral index is $\alpha =
-0.92\pm 0.08$ ($S\propto \nu^{\alpha}$) (Harnett~\cite{Harnett84});
this indicates that the dominant radio continuum emission
mechanism is synchrotron radiation rather than thermal processes.


\section{Observations and data reduction}

In the continuum observation mode at the ATCA, data are acquired in
two independent linear polarizations. The standard ATCA primary flux
density calibrator J1934-638 was observed at the start and end of each
synthesis observation and secondary (phase) calibrators were observed
regularly to give good parallactic angle coverage. In
Table~\ref{tab:obs} we present particulars of the $\lambda$6~cm ATCA
observations of NGC\,2442.

\begin{table}[ht]
  \begin{center}
    \caption{Observational data at $\lambda$6~cm (5170~MHz)
     for NGC\,2442}
    \label{tab:obs}
    \begin{tabular}{lcc}
      \hline\hline
      \noalign{\smallskip}
      Date & Configuration & Time on source (h)\\
      \noalign{\smallskip}
      \hline
      \noalign{\smallskip}
      1996 Jan 19 & 750C & 3\\
      1996 Feb 9 + 11 & 750B& 4\\
      1996 Nov 2 & 750A & 10\\
      1998 Mar 25 & 375 & 9\\
      2000 Dec 31 & 750C & 11\\
      \noalign{\smallskip} \hline \
    \end{tabular}
  \end{center}
\end{table}

The \emph{MIRIAD} data reduction package (Sault \&
Killeen~\cite{Sault98}) was used for data reduction. Each data set of
total bandwidth 128 MHz and 32 channels was edited and calibrated
separately, data from all observations were then combined and reduced
in the standard way (Ehle et al.~\cite{Ehle96}; Sault \&
Killeen~\cite{Sault98}). Images shown here were made excluding 6-km
baseline data.

Stokes Q and U parameter maps were used with the \emph{MIRIAD}
polarization software to generate images of linearly polarized
intensity (PI, corrected for positive bias) and position angle (PA).
The final maps of total and polarized intensities at 10 $\arcsec$
resolution are shown in Figs.~\ref{fig:i6} and \ref{fig:pi6}.

We also obtained data for NGC\,2442 at \wav{13} from the ATCA.
Details are given in Beck et al. (\cite{Beck02}). The map of total
intensity at \wav{13} suffers severely from missing short spacings as
only the 1.5~km array configuration was used, whereas the \wav{6}
observations utilised both the 750~m and 375~m arrays.  Polarized
intensity is weak at \wav{13}; significant signal-to-noise ratios are
reached only after smoothing to $45\arcsec$ (Fig.~\ref{fig:pi13}).

With only one total intensity map at high angular resolution, we were
unable to separate the thermal and non-thermal (synchrotron) emission
components. The typical average thermal fraction in spiral galaxies at
$\lambda$6~cm is $\simeq20\%$ (e.g. Niklas et al.~\cite{Niklas97}),
but may locally reach $\simeq50\%$ in star-forming regions.

For comparison with the radio images, we made a continuum subtracted
\ion{H}{$\alpha$} map (see Fig.~\ref{fig:pi6}) using observations of
NGC~2442 by Ryder \& Dopita (\cite{Ryder93}), smoothed to a
resolution of $10\arcsec$.

\section{Results} \label{Res}

\begin{figure}[ht]
\resizebox{\hsize}{!}{\includegraphics[]{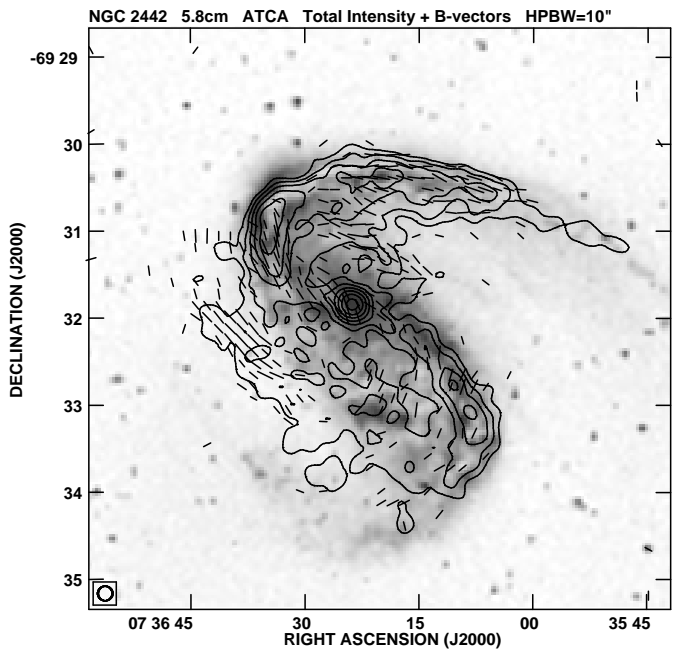}}
\caption{NGC\,2442: Total $\lambda$6~cm emission shown as contours
  overlaid on a greyscale image from the DSS. The half-power width
  (HPBW) of the synthesized beam is $10\arcsec$.  Contours are 1, 2,
  4, 8, 16, 32, 64 $\times$ the basic contour level of $10^{-1}$
  mJy/beam. Straight lines are E-vectors rotated by $90\degr$, not
  corrected for Faraday rotation, lengths scale with polarized
  emission and $1\arcsec$ represents $10^{-2}$ mJy/beam.  }
\label{fig:i6}            
\end{figure}

\begin{figure}[ht]
\resizebox{\hsize}{!}{\includegraphics[]{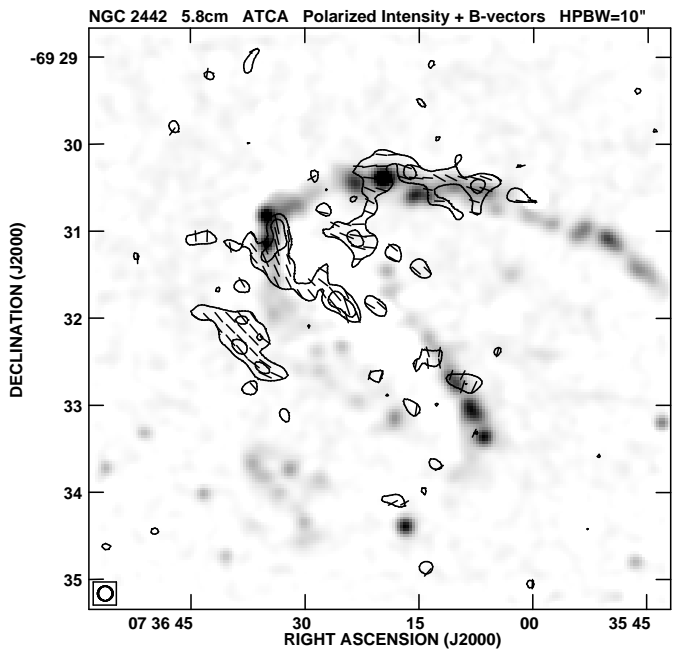}}
\caption{NGC 2442: Polarized emission at $\lambda$6~cm shown
  as contours overlaid on a greyscale image of H$\alpha$ emission.
  The half-power width (HPBW) of the synthesized radio beam is
  $10\arcsec$ and the H$\alpha$ image has been smoothed to this
  resolution. Contours are 3, 5 $\times$ the basic contour level of
  25~$\mu$Jy/beam. Straight lines are E-vectors rotated by $90\degr$,
  not corrected for Faraday rotation, lengths vary with polarized
  emission and $1\arcsec$ represents $10 \mu$Jy/beam}
\label{fig:pi6}            
\end{figure}

In Figure~\ref{fig:i6} we present the total $\lambda$6~cm continuum
emission from NGC\,2442 as a contour plot with B-vectors (the
observed orientation of the electric polarization vector rotated by
$90\degr$, without correction for Faraday rotation), their lengths
proportional to the polarized intensity, overlaid on a Digital Sky
Survey (DSS) image. The restoring HPBW is $10\arcsec$.  The integrated
total flux density within a radius of $3\arcmin$ is $S=123\pm 12$~mJy
and the integrated polarized intensity (Fig.~\ref{fig:pi6}) is $20\pm
5$~mJy, giving an overall degree of polarization of $16\pm 4$~\%. Both
images have an rms noise $25\,\mu$Jy/beam. The total flux density is
significantly higher than the $74\pm 4$~mJy reported by Beck et
al. (\cite{Beck02}), probably due to a baseline error in the earlier
map. Using a \wav{6} flux density of $80\pm 10$~mJy, Harnett
(\cite{Harnett84}) derived a spectral index of $\alpha=-0.92$
($S\propto\nu^{\alpha}$) for NGC\,2442 -- unusually steep for spiral
galaxies (Condon~\cite{Condon92}). The higher flux density determined
using our new \wav{6} map is compatible with a flatter spectral index
of $\alpha \simeq -0.75$, more typical for spiral galaxies.

\begin{figure}[ht]
\resizebox{\hsize}{!}{\includegraphics[]{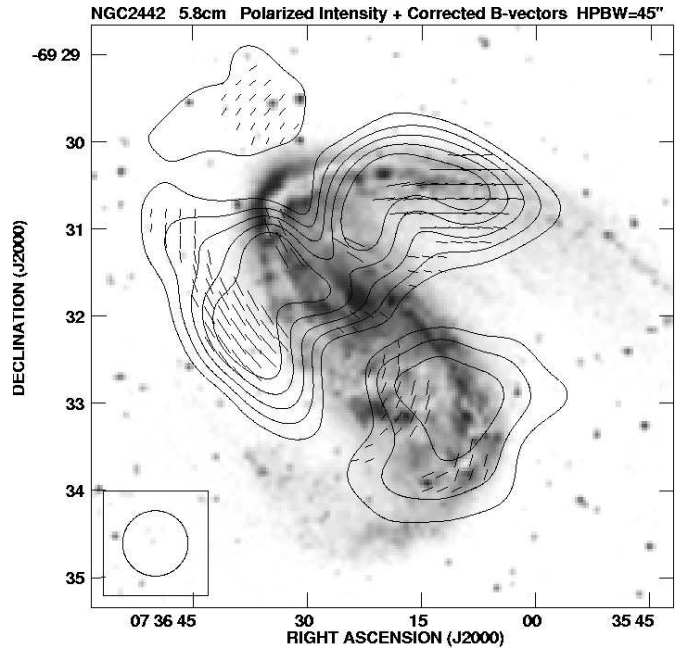}}
\caption{NGC 2442: Polarized emission at $\lambda$6~cm shown
  as contours overlaid on a greyscale image from the DSS.  The
  half-power beam width (HPBW) is $45\arcsec$.  Contours are 1, 1.5,
  2, 2.5, 3, 3.5 $\times$ the basic contour level of
  4~$\times$~$10^{-2}$ mJy/beam. Straight lines are E-vectors rotated
  by $90\degr$ and corrected for Faraday rotation, thus showing the
  orientation of the plane-of-sky regular component of the magnetic
  field in NGC 2442. The lengths of the ``vectors'' vary with
  polarized emission and $1\arcsec$ represents $63 \mu$Jy/beam.}
\label{fig:pi6:45}            
\end{figure}

\begin{figure}[ht]
  \resizebox{\hsize}{!}{\includegraphics[]{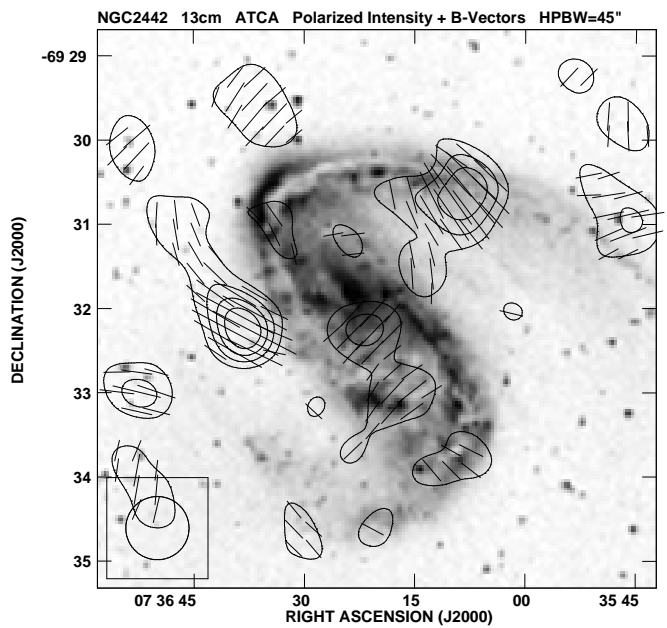}}
\caption{NGC 2442: Polarized emission at $\lambda{13}$~cm emission
  shown as contours overlaid on a greyscale image from the DSS.  The
  half-power beam width (HPBW) is $45\arcsec$.  Contours are 1, 2, 4,
  8, 12, 16, 32, 64 $\times$ the basic contour level of
  7~$\times$~$10^{-2}$ mJy/beam. Straight lines are E-vectors rotated
  by $90\degr$, lengths vary with polarized emission and $10\arcsec$
  represents $5\times10^{-2}$ mJy/beam }
\label{fig:pi13}               
\end{figure}

The total intensity arms are noticeably wider than the H$\alpha$ arms
shown in greyscale in Fig.~\ref{fig:pi6}, a clear indication that much
of the radio continuum emission is non-thermal and that cosmic ray
electrons diffuse a long way from their presumed sources in supernova
remnants. In general, the structure in total intensity mimics the
large-scale optical morphology to at least the $40\,\mu$Jy/beam level.
Emission from the deformed spiral arms and the nuclear region, where
the peak intensity of 8.8~mJy/beam occurs, are all apparent in
Figure~\ref{fig:i6}.  The small optical bar is barely visible.  There
is no circum-nuclear ring in radio continuum, even with our best
angular resolution of 10\arcsec. The most striking aspect of the
emission is the steep gradient at the outer edge of the prominent
northern arm.  In the following, we will call the extended northern
arm the ``peninsula''.  Here the emission runs parallel and close to
the well-delineated dust lane.  There is also a local maximum of 2.0
mJy/beam at RA $07$h~$36$m~$20$s, DEC
$-69\degr$~$30\arcmin$~$30\arcsec$, that appears to coincide with two
optically prominent $\ion{H}{ii}$ complexes. Diffuse emission also
extends to the west and south in the region enclosed by the northern
arm, where no optical counterpart is apparent.  We will call this area
around RA $07$h~$36$m~$20$s, DEC $-69\degr$~$31\arcmin$ the ``bay'' in
our discussion below.

In regions where spiral arms turn suddenly, theoretical models predict
enhanced turbulence and velocity shear (Roberts et
al.~\cite{Roberts79}; Athanassoula~\cite{Athanassoula92}; Lindblad et
al.~\cite{Lindblad96}).  Increased star formation and tangling of
magnetic field lines are expected to result in an increase in total
synchrotron emission, but also lower the degree of polarization. We
see some evidence of this at the $\sim90\degr$ bend in the northern
arm, where strong H$\alpha$ emission and strong total radio emission
of up to 2 mJy/beam is detected, whilst the linearly polarized
component rapidly disappears (see Figs.~\ref{fig:i6} \&
\ref{fig:pi6}).

South of the nucleus, the $\lambda$6~cm image (Fig.~\ref{fig:i6})
reveals extended, smooth emission from the less well-defined southern
arm, with local maxima and a sharp western edge.  To the southeast,
the emission follows the faint southern spiral arm after it has
turned. There is a local maximum near to the turn itself. Only small
patches of polarized emission are detected near the southern arm
(Fig.~\ref{fig:pi6}).

No enhanced emission was detected from the position of SN1999ga -- RA
$7$h~$36$m~$17$s, DEC $-69\degr$~$33\arcmin$~$22\arcsec$ (Woodings et
al. \cite{Woodings99}).

First seen in the MOST images (Harnett~\cite{Harnett84}), significant
large-scale non-thermal emission to the east is visible in
Fig.~\ref{fig:i6}, without any optical or H$\alpha$ counterpart. The
extremely high degree of polarization makes this feature prominent in
polarized intensity (Fig.~\ref{fig:pi6}). In the following, we will
refer to this feature as the ``island'' (see
Sect.~\ref{subsec:island}). Note that none of the three regions -- the
peninsula, the bay and the island -- are artifacts of resolution
(compare Figs.~\ref{fig:pi6} and \ref{fig:pi6:45}).

\begin{figure}[ht]
\resizebox{\hsize}{!}{\includegraphics[]{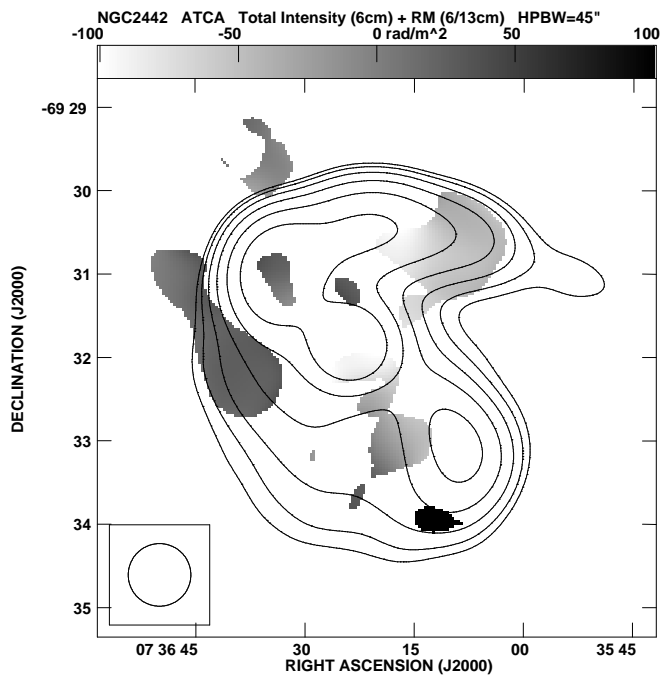}}
\caption{NGC 2442: Rotation measures between $\lambda{13}$~cm and
  $\lambda{6}$~cm shown as a greyscale image with contours
  representing the total $\lambda{6}$~cm emission with a HPBW of
  $45\arcsec$. Contour levels are 1, 2, 4, 8, 12, 16, 32 $\times$ the
  basic contour level of 2~$\times$~$10^{-1}$ mJy/beam}
\label{fig:rm}            
\end{figure}

The orientations of the B-vectors and the extent of the polarized
emission are presented in Fig.~\ref{fig:pi6} as overlays on an
H$\alpha$ image.  The large-scale regular magnetic field is patchy
with maximum extent along the peninsula; note that the ridge of
polarization along the peninsula is clearly shifted to the outside of
the H$\alpha$ arm. From the regions of turbulence, where the arms turn
suddenly and strong wavelength-independent depolarization is expected
to occur, the polarized emission from the northern arm abruptly
weakens. No polarized emission was detected from the turn in the
southern arm. Generally in the southern arm, only weak polarized
emission is evident. We detected some diffuse polarized emission after
smoothing (see Figs.~\ref{fig:pi6:45} \&
\ref{fig:pi13}). Figure~\ref{fig:pi6:45} shows the orientation of the
plane-of-sky component of the regular magnetic field in NGC~2442,
obtained by rotating \wav{6} E-vectors by $90\degr$ and correcting for
the effects of Faraday rotation (Fig.~\ref{fig:rm}). The regular field
is thought to lie predominantly in plane of most disc galaxies (Beck
et al. \cite{Beck96}), in which case Fig.~\ref{fig:pi6:45} shows the
intrinsic orientation of the regular magnetic field in NGC~2442. Note
that Fig.~\ref{fig:pi6:45} has been generated by smoothing the \wav{6}
polarized emission (displayed in Fig.~\ref{fig:pi6}) by a factor of
$4.5$ to match the angular resolution at \wav{13}
(Fig.~\ref{fig:pi13}). The corrected vectors are plotted only where
polarized emission at both wavelengths is above the $3\times$ the
noise level.

The best-aligned and strongest regular fields are located along the
inner northern arm, the peninsula, and at the island and the bay. In
the peninsula, the vectors are coherent over $\sim 90\arcsec\simeq
7$~kpc and are generally oriented parallel to the dust lane, which
may be a sign of a large-scale shock (see Sect.~\ref{subsubsec:sln}).

Faraday rotation measures (RM) between \wav{6} and \wav{13} are shown
in Fig.~\ref{fig:rm} at a resolution of $45\arcsec$ (no correction for
Faraday rotation due to the Milky Way foreground has been applied).
Two large, coherent patches of RM are visible: at the position of the
island, where $18<RM<27\radm$, and at the peninsula, where
$-60<RM<-35\radm$. The smooth variation and constant sign of the RMs
in these two patches means that the magnetic field has a coherent,
uni-directional regular component (see e.g. Beck et al. \cite{Beck96a}
for more details on Faraday rotation and galactic magnetic fields).

The different signs of RM in the island and peninsula seem to indicate
that the regular field is pointing in different directions.  However,
the island is located near the minor axis (which has a position angle
of $130\degr$) of the projected galaxy plane where any plane-parallel,
predominantly azimuthal, regular field has only a small line-of-sight
component. Hence, we suspect that the positive RM in the island is due
to the foreground plasma in the Milky Way; indeed, values around
$+20\radm$ are consistent with the foreground RM expected near
Galactic coordinates $l\simeq280\degr, b\simeq-20\degr$ (Han et al.\
\cite{Han97}). If so, the measured RMs from the peninsula are also
affected by the foreground RM and the intrinsic values in fact are
$-90\lesssim$RM$\lesssim-55\radm$. Assuming trailing spiral arms, we
learn from the velocity field (Bajaja et al. \cite{Bajaja99}) that the
northwestern side with the peninsula is the far side of the
galaxy. This means that the radial component of the regular
magnetic field is directed away from the centre of the galaxy (if we
assume that the regular field lies predominantly in the plane of the
galaxy), in contrast to the majority of galaxies studied so far
(Krause \& Beck \cite{Krause98}).


\section{Magnetic field properties}
\label{sec:B}

\subsection{Asymmetry of \wav{6} emission}
\label{subsec:asym}

To investigate the symmetry of the emission in NGC\,2442, we have
averaged the flux density in two regions separated by the minor
axis. The intense emission all along the northern arm results in a
strong asymmetry in total intensity.  We find the north/south ratio of
polarized $\lambda 6$~cm flux densities in the two regions to be,
rather surprisingly, only 1.6 compared to 3.4 for the total
emission. While the polarized emission shows distinct features only in
the northern half, there is weak diffuse polarized emission in the
southern half (Fig.~\ref{fig:pi6:45}).

\subsection {Local magnetic field strengths}
\label{subsec:magB}

\begin{table*}[ht]
  \begin{center}
    \caption{Total power, polarized intensity and equipartition magnetic field 
     strengths in NGC 2442}
    \label{tab:B}
    \begin{tabular}{lcccc}
      \hline\hline
      \noalign{\smallskip}
      & Island & Northern arm at turn & Peninsula & Bay\\
      \noalign{\smallskip}
      \hline
      \noalign{\smallskip}
      Mean total intensity ($\mu$Jy/beam) &
	 $198\pm{23}$ & $1606\pm{22}$ & $593\pm{22}$ & $271\pm{17}$ \\
      Mean non-thermal intensity ($\mu$Jy/beam)   &
	 $198\pm{23}$ & $1166\pm{22}$ & $415\pm{22}$ & $68\pm{17}$ \\
      Mean polarized intensity ($\mu$Jy/beam)    &
	 $102\pm{23}$ & $87\pm{22}$   &  $77\pm{22}$ & $101\pm{17}$ \\
      Percentage polarized emission              &
	 $52\pm{13}\%$  & $7\pm{2}\%$    &  $19\pm{5}\%$ & $36\pm{9}\%$ \\
      Total magnetic field strength ($\mu$G)     &
	 $16\pm{5}$  & $26\pm{8}$   &  $20\pm{6}$ & $16\pm{5}$ \\
      Regular magnetic field strength ($\mu$G)   &
	 $13\pm{4}$  & $7\pm{2}$    &   $9\pm{3}$ & $10\pm{3}$ \\
      \noalign{\smallskip} \hline \
    \end{tabular}\\
  \end{center}
Notes: errors in radio emission intensities are the rms noise weighted
by the area over which the average emission was calculated; errors in
equipartition magnetic field strengths are estimated to be $30$\%
(see the text for details).
\end{table*}

Table~\ref{tab:B} lists the mean total, non-thermal and polarized
intensity as well as the magnetic field properties of the island,
inner northern arm at the turning point, peninsula and bay. The total
magnetic field strength $\Bt$ is estimated assuming energy
equipartition between cosmic ray particles and the magnetic field
(e.g.  Longair~\cite{Longair94}, p.292). The strength of the regular
magnetic field $\Br$ is derived from the degree of polarization.  The
following assumptions are used: the depth of the emitting layer is
1~kpc, the ratio of the energies present in relativistic protons to
electrons is 100, the low energy cutoff for cosmic ray protons is 300
MeV, and the synchrotron spectral index is $\alpha\simeq-0.9$
(Harnett~\cite{Harnett84}).  We integrate over a fixed energy interval
in the cosmic-ray spectrum to obtain the total energy density of the
cosmic rays independent of the magnetic field strength (see
Beck~\cite{Beck00}). The thermal fraction in the spiral arms and the
bay is assumed to be 30\%, but the emission from the island at
$\lambda$6~cm is taken to be completely non-thermal, as indicated by
the high percentage of polarized emission and absence of H$\alpha$
emission.  Most of these assumptions are canonical, based on estimates
for the Milky Way and a few other galaxies, and are clearly subject to
large errors. The derived magnetic field strength is proportional to
the power $(3-\alpha)^{-1}\simeq 1/4$, so even large uncertainties
cause only small errors in the deduced field strength; we estimate
typical errors of $\pm 30\%$.

The total magnetic field strength is high in the northern spiral arm,
probably as a result of stronger driving of interstellar turbulence
by, for example, star formation and gravitational instabilities. The
very highly polarized emission in the island and bay means that the
field here is mostly regular. If the equipartition assumption is
valid, the regular magnetic field strength estimates of $\Br\simeq
12\mkG$ are very high, comparable to $\Br$ in the magnetic arms of
NGC\,6946 (Beck~\cite{Beck96}).

\subsection{Comparison of total and polarized radio emission at \wav{6}}
\label{subsec:slices}

\begin{figure}[ht]
\resizebox{\hsize}{!}{\includegraphics{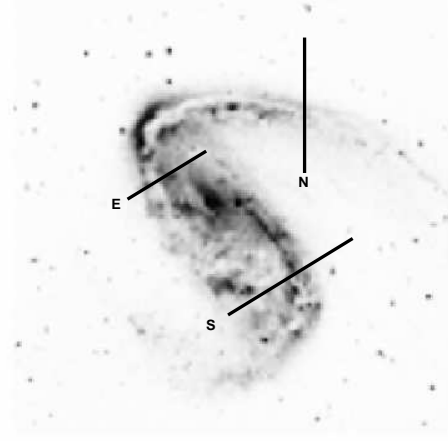}}
\caption{NGC 2442: DSS image of NGC 2442 showing positions of slices
  discussed in the text. The position of the label on each slice shows the
  left end of the x-axis in Figs.~\ref{fig:sln}, \ref{fig:sle} \&
  \ref{fig:sls}}
\label{fig:slices}            
\end{figure}

In Figure~\ref{fig:slices}, we show the positions of slices through
the arms of NGC\,2442, that we used to compare the emission in total
power, polarized emission, infrared K-band (Ryder unpublished)
and H$\alpha$. We found that the K-band and H$\alpha$ have similar
normalized profiles and therefore we only present H$\alpha$ data. For
each slice, the position of the label denotes the left end of the
x-axis in Figs.~\ref{fig:sln},
\ref{fig:sle} \& \ref{fig:sls}.

Before making the slices 'N' and 'S' we smoothed the radio and
H$\alpha$ maps at original resolution ($10\arcsec\times 10\arcsec$)
with a highly elliptical, $10\arcsec\times 30\arcsec$ beam, with the
major axis of the smoothing beam oriented parallel to the local
direction of the arm. This has a similar effect to averaging several
neighbouring slices, and so reduces the risk of developing an
interpretation of the emission profiles based on an atypical slice. As
a further check we made more slices through the original maps, not
shown here, and found that the cross-section obtained following
smoothing with an elliptical beam is a representative, average
profile. The smoothing also increases the signal to noise ratio of any
diffuse emission, especially in the weak southern arm, without losing
resolution in the direction perpendicular to the arm.  Slice 'E' is
based on the original unsmoothed map.

\subsubsection{The peninsula: Slice N}
\label{subsubsec:sln}

\begin{figure}[ht]  
\resizebox{\hsize}{!}{\includegraphics[]{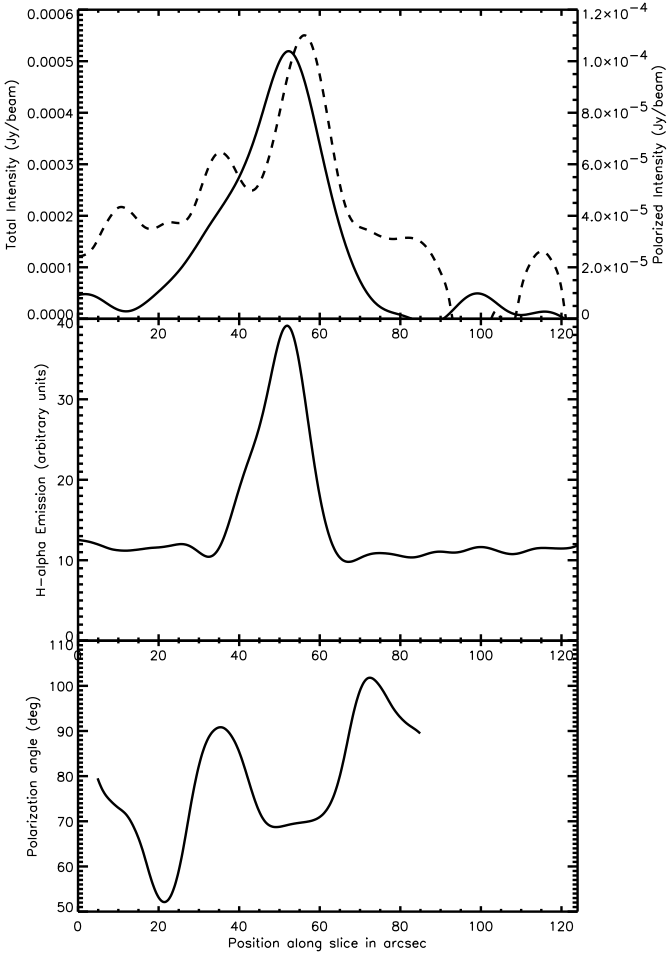}}
\caption{Slice across the bay and the peninsula
  (position N in Fig.~\ref{fig:slices}).  South is to the left.
  \emph{Upper panel :} total (solid line) and polarized (dashed line)
  \wav{6} emission, both smoothed to a beam of
  $10\arcsec\times30\arcsec$ oriented parallel to the spiral arm.
  \emph{Middle panel :} H$\alpha$ emission.  \emph{Bottom panel :}
  'B-vector' position angles}
\label{fig:sln}            
\end{figure}

Figure~\ref{fig:sln} shows that whilst the profile of the strongest
peak in polarized emission (PI) has an approximately Gaussian profile,
the total power (TP) profile is asymmetric, broader and is displaced
by $5\arcsec$ or approximately half a beam-width. As long as the signal
to noise ratio of the peaks are sufficiently high this displacement is
reliable \emph{even though the shift is less than the beam-size
$\theta$}. To illustrate this, assume that the signal at a peak has a
Gaussian profile $S_1=S_0 exp(-(x-x^{\prime})^2/(\theta)^2)$ with
amplitude $S_0$, centred on $x^{\prime}$ and that this signal is
combined with random noise that serves to produce a second maximum
within the beam $S_2=N_0 \delta (x-x^{\prime})/\theta$, of amplitude
$N_0$ shifted by a random factor $\delta$ from the true
peak. Expanding $S_1$, adding $S_1$ to $S_2$ and differentiating gives
an estimate for the positional accuracy of the peak $\Delta x$ in
terms of the beam-size $\theta$ and the signal to noise ratio $S_0/N_0$
\begin{equation}
	\label{eq:position}
	\Delta x \simeq \frac{N_0 \delta \theta}{2
	S_0} \sim \frac{\theta}{2 S_0/N_0},
\end{equation}
where we conservatively assume that the noise has a maximum right at
the edge of the beam (i.e. $\delta=1$). With $TP= 500\mu \mathrm{Jy}$,
$\sigma_{TP}=25\mu\mathrm{Jy}$, $PI=110\mu\mathrm{Jy}$ and
$\sigma_{PI}=25\mu\mathrm{Jy}$ at the peak in slice N, using
Eq.~\ref{eq:position} we expect the error in position of the TP peak
to be $\pm 0.5\arcsec$ and that in PI to be $\pm 2\arcsec$.

Given the distance of 15.5~Mpc, the emission maxima are separated by
$\sim 400$~pc.  The peak of the H$\alpha$ and TP profiles coincide
exactly, but the former is narrower (Fig.~\ref{fig:sln}, middle
panel). We conclude that along this slice, the strongest total
emission (comprising thermal + non-thermal emission) coincides with a
region of significant star formation, as indicated by the H$\alpha$
profile.  This coincidence is in obvious contrast to the inner
northern arm (Fig.~\ref{fig:sle}) and indicates that the peninsula is
not subject to density-wave compression, where a shift between the
peaks of thermal and non-thermal emission is expected
(Sect.~\ref{subsubsec:sle}).

The polarized emission and therefore the regular magnetic field is
strongest toward the $\it{outer}$ edge of the peninsula, i.e. to the
north of the star-forming region (Fig.~\ref{fig:sln}).  This shift
between the total and polarized emission could be due to Faraday
depolarization within the northern arm. Intrinsic Faraday rotation
measures between $\lambda$6~cm and $\lambda$13~cm are $\simeq
-70~$rad\,m$^{-2}$ in the peninsula (Sect.~\ref{Res}). Another
estimate of the amount of Faraday rotation is based on the variation
of polarization angles (Fig.~\ref{fig:sln}, bottom panel); the angles
in the spiral arm, around the peak in total emission, are smaller by
about $20\degr$ compared to the angles on either side, which
corresponds to a Faraday rotation of $\simeq -90$~rad/m$^2$ and to
very weak depolarization by differential Faraday rotation of 0.92 (see
Sokoloff et al \cite{Sokoloff98} for a detailed discussion of Faraday
depolarization).  This is insufficient to account for the difference
in degree of polarization ($p=PI/TP$) between the peak in TP, where
$p=18\%$, and the peak in PI, where $p=24\%$.  Depolarization by
Faraday dispersion in the turbulent magnetic field of the peninsula is
another possibility.  However, if the shift in the PI peak is a
product solely of depolarization in a thermal plasma, we would expect
the PI profile to be symmetric about the H$\alpha$ profile. In other
words, there should be a PI minimum in the peninsula which is not the
case.

Our favoured explanation for the displacement of the regular field is
ram pressure exerted by the intergalactic medium.  The velocity field
is strongly distorted in the peninsula (Bajaja et al.
\cite{Bajaja99}). Ram pressure would increase the strength of the
magnetic field component perpendicular to the direction of the
pressure and tend to turn the B-vectors perpendicular to the streaming
velocity of the intergalactic medium, which we assume to be roughly in
the N-S direction.  This agrees reasonably well with the observed
position angles of the B-vectors in this region of about $90\degr$.

Highly polarized radio emission at the outer edges of spiral arms due
to ram pressure was also detected in NGC~2276 (Hummel \&
Beck~\cite{Hummel95}) and in the Virgo galaxy NGC~4254 (Soida et
al.~\cite{Soida96}).  Polarized emission is a very sensitive tracer of
such compression effects.

Tidal interaction may be another reason for the distorted peninsula
and its sharp outer edge, as demonstrated by N-particle models (e.g.
Combes et al.~\cite{Combes88}). Observations of \ion{H}{i} by Houghton
(\cite{Houghton98}) indicate gravitational disturbances in the
velocity fields of two galaxies near to NGC~2442, [VC2]073649.1-691448
and ESO059-G006. Ram pressure should compress \ion{H}{i} gas along the
outer edge of the peninsula, whilst no systematic shifts between gas,
magnetic fields and star-forming regions are expected in tidal arms.
Furthermore, tidal interaction can generate huge clouds of \ion{H}{i}
gas outside the spiral arms, but no such clouds were detected by
Houghton (\cite{Houghton98}).

Another explanation is that slice~N lies outside the co-rotation
radius so that the pattern speed is faster than the speed of the gas
orbits about the galactic centre.  Star formation is triggered by the
compression of the gas, where the arm catches up with it, i.e. to the
north of the northern spiral arm.  If the regular field were
subsequently randomized by turbulence associated with these processes,
we would expect the polarized emission to peak on the \emph{outside}
of the peninsula.  In strongly barred galaxies the co-rotation radius
lies at the ends of the bar. Although NGC~2442 has only a weak, short
bar, it seems possible that the large-scale distortion of the spiral
arms, particularly to the north-east, acts in the same way as a strong
bar (Houghton \cite{Houghton98}). To test the co-rotation assumption,
we require velocity field data and accurate modelling of the pattern
speed.

\subsubsection{The inner northern arm: Slice E}
\label{subsubsec:sle}

\begin{figure}[ht]  
\resizebox{\hsize}{!}{\includegraphics[]{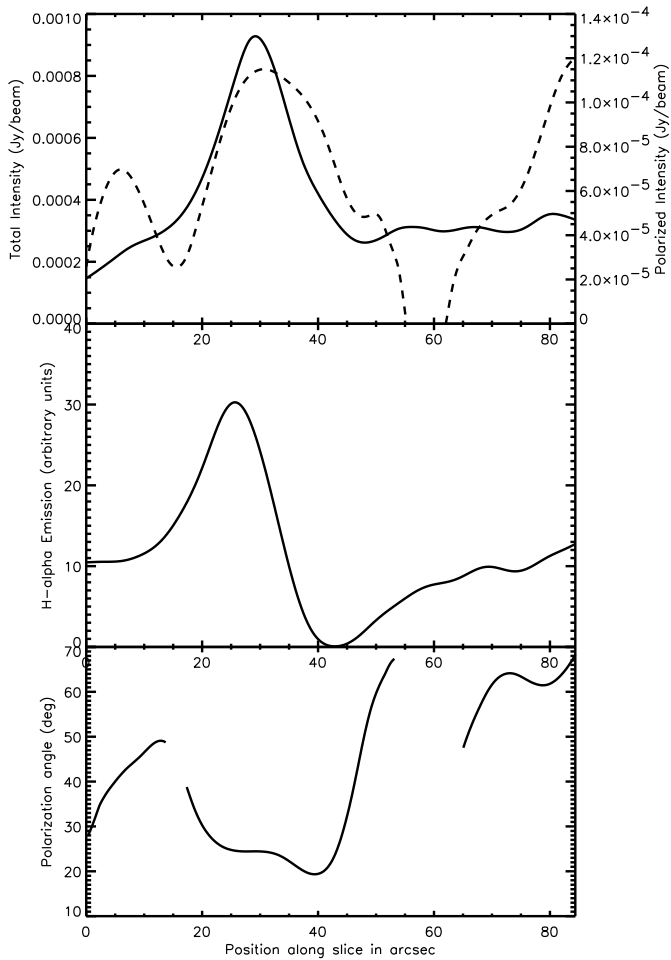}}
\caption{Slice across the northern arm (position E in Fig.~\ref{fig:slices}).
  East is to the left.  \emph{Upper panel :} total (solid line) and
  polarized (dashed line) \wav{6} emission, both at the original
  $10\arcsec\times10\arcsec$ resolution. \emph{Middle panel :}
  H$\alpha$ emission.  \emph{Bottom panel :} 'B-vector' position
  angles}
\label{fig:sle}            
\end{figure}

The upper panel of Fig.~\ref{fig:sle} displays the profile of TP and
PI in the direction E-W across the inner northern arm.  Note that the
map was not smoothed by an elliptical beam prior to making the slice;
this method is only appropriate when the Stokes parameters I, Q \& U
are roughly constant in the direction perpendicular to the slice.

Here there is a slight displacement between the peaks in TP and PI
emission.  However, the PI peak is significantly broader than that in
TP and the PI peak in slice 'N'. This indicates that the PI peak is
not generated by a ram pressure or a density wave shock. The
orientation of the B-vectors (see lower panel of
Figure~\ref{fig:sle}) jumps suddenly by $\sim 40\degr$, from $\simeq
20\degr$ in the arm to $\simeq 60\degr$ in the bay.

This situation is similar to that detected across the bar in NGC\,1097
(Beck et al.~\cite{Beck99}), where the regions of enhanced total
emission (TP) coincide with the dust lanes along the bar, while the
polarized emission is much broader and extends far into the upstream
region.  About $1$~kpc upstream of the dust lane, the position angles
of the B-vectors change abruptly, which is indicated by a
region of zero polarized intensity. Beck et al. (\cite{Beck99})
conclude that a shear shock, due to the bar's gravitational potential,
is responsible for both the deflection and amplification of the
magnetic field.

We see an indication of such a field deflection in NGC\,2442, but our
resolution does not match that of the NGC\,1097 image.  We propose
that a similar situation exists in NGC\,2442 due to the bar-like
distortion of the inner spiral arms.  This means that the magnetic
field along the ``bar'' aligns well with the local gas streamlines. We
predict a sharp change in the velocity field profile.

The peak in H$\alpha$ emission is shifted $\simeq 5\arcsec$ east, or
downstream, of the TP peak. This offset is consistent with the
interpretation that the gas at this radius rotates faster than the
spiral pattern speed. In this case, gas and magnetic field are
compressed when the gas enters the arm from the west, followed by star
formation further downstream.

\subsubsection{The southern arm: Slice S}
\label{subsubsec:sls}

\begin{figure}[ht]  
\resizebox{\hsize}{!}{\includegraphics[]{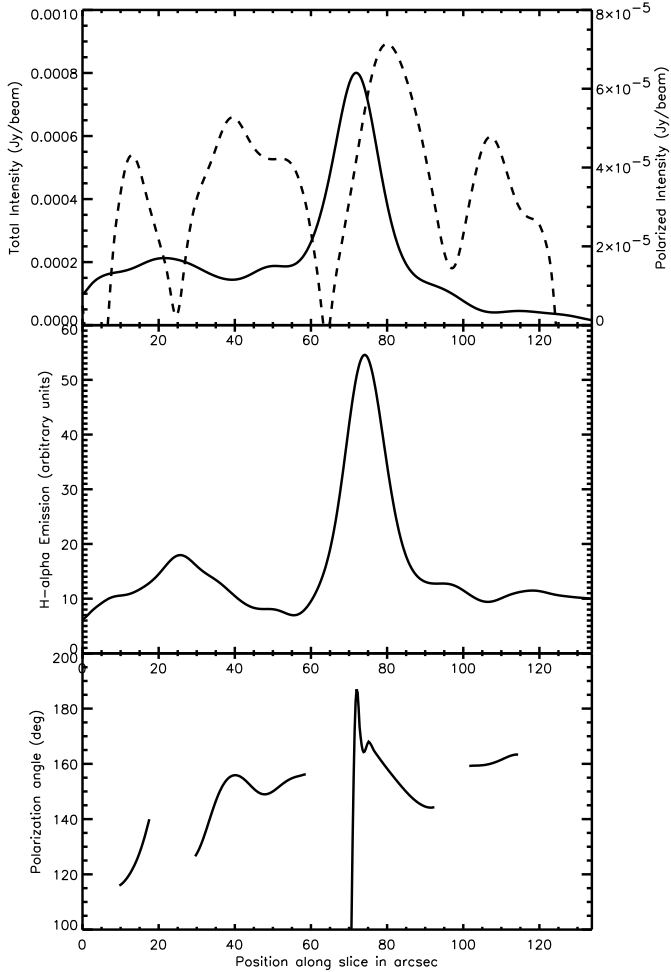}}
\caption{Slice across the southern arm (position S in Fig.~\ref{fig:slices}).
  East is to the left. \emph{Upper panel :} total (solid line) and
  polarized (dashed line) \wav{6} emission, both smoothed to a beam of
  $10\arcsec\times30\arcsec$ oriented parallel to the spiral arm.
  \emph{Middle panel :} H$\alpha$ emission.  \emph{Bottom panel :}
  'B-vector' position angles}
\label{fig:sls}            
\end{figure}

Figure~\ref{fig:sls} shows a slice through the southern spiral arm
marked S in Fig.~\ref{fig:slices}. In this region, the TP emission is
more extended and diffuse and the PI is patchy even after smoothing
with a beam extended parallel to the arm.  The position angles of the
B-vectors are roughly constant across the inner arm.  There is
an indication from our slice profile of the PI peak being shifted
toward the \emph{outer} edge, but the emission is very weak and patchy
and may not be shaped by large scale processes. Nevertheless, both the
sharp western edge of the spiral arm and the shift between the PI and
TP peaks in Fig.~\ref{fig:sls} resemble the conditions in the
peninsula (Fig.~\ref{fig:sln}) and indicate that ram pressure may also
act onto the southern arm.

\subsection {The island}
\label{subsec:island}

The island is displaced from the northern spiral arm by $\sim5\kpc$ --
measured in the galaxy's plane -- and is $\sim 5.5$~kpc long in the
high resolution map (Fig.~\ref{fig:pi6}). Here the continuum emission
is strongly polarized, up to 46\% of the total emission.  There is no
detected H$\alpha$ counterpart.

Without star-forming activity the island probably does not contain the
sources of the cosmic ray electrons producing synchrotron
emission. Hence, the relativistic particles have to travel about
5~kpc, from the nearest star-forming regions, without significant
energy losses. The timescale for synchrotron energy losses can be
approximated by
\begin{equation}
  \label{eq:eloss}
  t_{e}\simeq \frac{8.35 \times 10^9 \mathrm{yr}}
    {\sqrt{\nu\,(\mathrm{MHz})/16}\,\, \left(\Btpe (\mkG)\right)^{1.5}},
\end{equation}
(Lang \cite{Lang99}, Sect.~1.25) where $\Btpe\sim\Bt$ is the
plane-of-sky component of the total magnetic field. The streaming
velocity, and hence the diffusion timescale, for the cosmic ray
electrons is thought to depend on scattering of the particles by
plasma waves. Avoiding complications, such as the statistical nature
of interstellar turbulence and the magnetic field geometry, we make a
crude assumption that the electrons stream at the Alfv\'en speed
$v_\mathrm{A}=\Bt/2\sqrt{\pi\rho}$ where $\rho$ is the mass density of
the fully ionised gas (e.g. Longair~\cite{Longair94}, Sect.~20.4), and
so the diffusion timescale to reach the island is
\begin{equation}
  \label{eq:ediff}
  t_d \simeq \frac{L}{v_\mathrm{A}},
\end{equation}
where $L$ is the distance from the spiral arm to the island. For the
cosmic ray electrons to reach the island before radiating away
significant energy requires that $t_d<t_e$. Combining
Eq.~\ref{eq:eloss} and Eq.~\ref{eq:ediff} we obtain an upper limit on
the product of the thermal electron density of the ISM and the
magnetic field strength,
\begin{equation}
  \label{eq:neB}
  \sqrt{\Bt (\mkG) n_\mathrm{e} (\cmcube)}\lesssim 0.21,
\end{equation}
where we have assumed that the ISM is fully ionised hydrogen and
$n_\mathrm{e}$ is the thermal electron density. For a total magnetic
field strength between the northern arm and the island of $\Bt\simeq
15\mkG$ (similar to that in the bay), Eq.~\ref{eq:neB} gives an upper
limit of $n_\mathrm{e}\lesssim 0.003\cmcube$.  Weaker magnetic fields
or a denser ionised ISM would result in too slow an Alfv\'en speed for
the cosmic ray electrons to reach the island, from their presumed
sources near star forming regions.

If the cosmic ray electrons in the island are older than in the
northern arm, their spectrum should steepen with distance from the
places of origin. Unfortunately due to missing spacing problems at
\wav{13}, we have only single-frequency information on the total
intensity emission from NGC\,2442.  Our maps of polarized intensities
cannot be used for spectral index studies because of
wavelength-dependent depolarization mechanisms. Total intensity maps
at other frequencies are needed.

Could the island be due to compression of the magnetic field ?
Compared to the compression feature in the peninsula, the island is
too broad and, more importantly, on the ``lee'' side of the galaxy
where no enhanced compression by ram pressure is expected or evident.

Could the island with its well-aligned vectors be the signature of a
``magnetic arm'' of the type seen between the optical arms of galaxies
such as NGC\,6946 (Beck \& Hoernes~\cite{Beck96}; Frick et
al.~\cite{Frick00})?  In Table~\ref{tab:B} the estimated regular
magnetic field strength is comparable to that of the magnetic arms in
NGC\,6946. The images at $45\arcsec$ resolution
(Figs.~\ref{fig:pi6:45} and \ref{fig:pi13}) show that the island
extends to the north, indeed approaching the shape of a magnetic arm.
Further, the B-vectors in the smoothed \wav{6} map
(Fig.~\ref{fig:pi6:45}) are oriented parallel to the inner northern
spiral arm, even though they are $\sim 5\kpc$ downstream of the arm.
This alignment can hardly be pure coincidence.

Another explanation, in view of the results presented above, is that
the regular field is oriented with the B-vectors parallel to
shear motion in the tidal gas flow of the disturbed galaxy. The island
might then represent a region of strong shear which enhances the
magnetic field.  The published velocity field (Bajaja et
al. \cite{Bajaja99}) does not include the island region. Better data
on Faraday rotation could test this possibility: sheared random
magnetic fields generate strong polarized emission but little Faraday
rotation whereas dynamo generated magnetic arms are strong in both
quantities.

\section{Conclusions}
\label{sec:conc}

\subsection{Summary of the observations}

Our ATCA maps of the southern peculiar galaxy NGC 2442 in total and
polarized intensity at $\lambda$6~cm and in polarized intensity at
$\lambda$13~cm, in comparison with our new H$\alpha$ map, reveal the
following features :

-- The spiral arms in total intensity are strongly deformed, like a
large oval distortion, similar to the optical arms. The small optical
bar is barely visible in radio continuum.

-- The steep gradients in the radio emission at the northern and
western edges indicate compression of the total magnetic field.

-- Along the spiral arms, the peaks in total intensity agree well with
those in H$\alpha$, but synchrotron emission makes the radio arm --
and hence the magnetic field and cosmic ray distributions -- broader
than the H$\alpha$ arm.

-- The regular magnetic field has an unusual morphology and is
concentrated along the outer northern arm (the ``peninsula''), in the
region enclosed by the northern arm (the ``bay''), and in an
``island'' separated by about 5~kpc from the inner northern arm to the
east.

-- The regular magnetic field in the peninsula is systematically
shifted outwards with respect to the spiral arms seen in total
intensity and H$\alpha$ by $\sim 400$~pc. We interpret this as an
indication of ram pressure exerted by the intergalactic
medium. 

-- Across the inner northern arm, the orientation of the B-vectors
jumps suddenly by $\sim 40\degr$. This behaviour is similar to a shear
shock in the barred galaxy NGC\,1097 (Beck et al. \cite{Beck99}), and
indicates the presence of a major oval distortion of the gravitational
field in NGC\,2442.

-- The highly polarized ``island'' is probably the peak of a magnetic
arm between the two spiral arms, similar to the magnetic arms
previously found in normal spiral galaxies.  If magnetic arms are
phase-shifted images of the preceding optical arm, the strong
polarized emission from this feature may reflect a phase of strong
star formation about $10^8$y ago, when the southern arm was exposed to
ram pressure.

-- The total magnetic field strength reaches $25\mu$G in the inner
northern arm. The regular field strength in the island is
$\simeq12\mu$G. Both values are higher than is usual for spiral
galaxies.

\subsection{Some astrophysical implications}

Our observations of NGC\,2442 show that polarized radio emission is an
excellent tracer of field compression due to interaction with the
inter-galactic medium.  Similar signatures of interaction have been
identified in two other galaxies, NGC\,2276 (Hummel \& Beck
\cite{Hummel95}) and NGC\,4254 in the Virgo Cluster (Soida et
al. \cite{Soida96}).

The importance of the magnetic field to the distribution and dynamics
of the ISM can be gauged by comparing the magnetic energy density to
the energy density of the gas. In Section~\ref{subsec:magB} we derived
estimates of the total magnetic field strength of $16\mu$G in the
``island'' inter-arm region and $26\mu$G in the northern spiral arm,
assuming equipartition between cosmic ray and magnetic field energy
densities. These field strengths translate into energy densities of
$U_\mathrm{B}\simeq$ 1 \& 3$\times 10^{-11}\erg\cmcube$, respectively.
The magnetic energy density in the arms is equivalent to that of
neutral gas clouds with turbulent velocity $10\kms$ and number density
$35\cmcube$; this density is of the same order as the cold neutral
medium in the Milky Way.  (We are not aware of any measurements of
local, neutral gas densities in the spiral arms of NGC\,2442.) In the
inner disc of NGC\,6946 Beck (\cite{Beck04}) determined the average
density of the neutral gas to be $20\cmcube$ and the turbulent energy
density to be in equipartition with that of the magnetic field.

The importance of the magnetic field in the warm diffuse gas can be
better estimated. Bajaja et al. (\cite{Bajaja99}) found a temperature
of $T\simeq 6500$\,K and volume density of $n_\mathrm{e}\simeq
10\cmcube$ for a typical \ion{H}{ii} region in the northern arm of
NGC\,2442. The thermal energy density inside such a region is
$U_\mathrm{th}\simeq 1\times 10^{-11}\erg\cmcube$, giving a plasma
$\beta=U_\mathrm{th}/U_\mathrm{B}\simeq 0.3$; the magnetic field may
be important for the dynamics and morphology of \ion{H}{ii} regions.
A similarly low value for $\beta$ was derived for NGC\,6946 (Beck
\cite{Beck04}).

As the cosmic-ray electrons emitting in the island probably originate
from the spiral arm and lose some fraction of their energy during
their diffusion, the equipartition assumption gives a lower limit of
the field strength in the inter-arm region. Such a strong field
outside star-forming regions is exceptional and poses the question of
the dynamical importance of the magnetic field. Little diffuse
H$\alpha$ emission is observed (Fig.~\ref{fig:pi6}; Mihos \& Bothun
\cite{Mihos97}). We estimated an upper limit of $n_\mathrm{e}\simeq
0.003\cmcube$ in the ``island'' inter-arm region in
Sect.~\ref{subsec:island}, using the energy loss and diffusion
timescales of cosmic ray electrons. Taking a canonical value of
$T=10^{4}$\,K for the temperature of the diffuse gas we obtain
$\beta\simeq 10^{-4}$: the magnetic field will be strongly dominant in
the local dynamics of the inter-arm gas. However, note that the energy
density associated with galactic rotation is several orders of
magnitude higher than that of the magnetic field.

The radio emission from the island is highly polarized and so is
clearly synchrotron radiation. But where do the cosmic ray electrons
come from? There is no emission in H$\alpha$ from this region,
indicating no star formation. The timescale for synchrotron energy
loss at \wav{6} in a $15\mkG$ field is about $10^{7}$ years
(Eq.~\ref{eq:eloss}) whereas a half rotation time -- when this region
would have been part of a spiral arm -- is about $10^{8}$ years (using
$\simeq250$km/s at 8~kpc radius, see Bajaja et
al. \cite{Bajaja99}). The cosmic rays most probably originate in the
spiral arms and travel at least 5~kpc in their lifetime. With an
independent measure of the gas density, the quality of the assumption
that the diffusion timescale of cosmic rays is proportional to the
reciprocal of the Alfv\'en speed (Eq.~\ref{eq:ediff}) could be tested
against observations.

The origin of the $\sim 5\kpc$ long, ordered magnetic field in the
island is not at all clear. Whereas the regular fields in the northern
arm show strong signs that they are the result of compression of a
tangled field, the inter-arm regular fields do not. Plausible
mechanisms for producing strong, well ordered magnetic fields away
from the spiral arms are enhanced dynamo action and shear arising from
differential rotation of the gas disc. High-resolution observations of
the velocity field, dynamo models and MHD models of the interaction
are required to test this.

The images of polarized emission at $45\arcsec$ resolution
(Figs.~\ref{fig:pi6:45} and \ref{fig:pi13}) indicate that the island
is the peak of a ``magnetic arm'', similar to that found in NGC\,6946
(Beck \& Hoernes~\cite{Beck96}; Frick et al.~\cite{Frick00}).  The
magnetic arms in NGC\,6946 are phase-shifted images of the optical
spiral arms preceding in the sense of rotation (Frick et al.
\cite{Frick00}). Possible explanations are slow MHD waves (Fan \& Lou
\cite{Fan97}), where the waves in gas density and magnetic field are
phase shifted, or enhanced dynamo action in the inter-arm regions (Moss
\cite{Moss98}, Shukurov \cite{Shukurov98}, Rohde et al. \cite{Rohde99}).
In NGC\,2442 the preceding gas arm to the east has a very small
amplitude as seen in the \ion{H}{i} observations of Houghton
(\cite{Houghton98}). However, half a rotation ago ($\simeq 10^8$y, see
above), this arm was fully exposed to the ram pressure of the
intergalactic medium and, similar to the peninsula at present,
probably had a much stronger amplitude in gas density. Star formation
has ceased since then, but the magnetic field wave has survived.
Hence, magnetic arms seem to preserve memory of the past phase of
active star formation. Polarization observations with higher
sensitivity and Faraday rotation measures at a higher resolution are
needed to discriminate between the possible origins if the ``island''.

\begin{acknowledgements}
  We thank Dmitry Sokoloff for helpful discussions and Elly
  M. Berkhuijsen and the referee for useful comments and carefully
  reading of the manuscript.

  JIH gratefully acknowledges the support and encouragement of the
  Alexander von Humboldt Society and the Max-Planck-Institut f\"ur
  Radioastronomie in Bonn, Germany, and in Australia of the University
  of Sydney and the Australia Telescope National Facility.

  ME is grateful to the ATNF for providing support and facilities. His
  work in Australia was funded through grant No. Eh 154/1-1 from the
  Deutsche Forschungsgemeinschaft.

  The Digitized Sky Survey was produced at the Space Telescope Science
  Institute under U.S Government grant NAG W-2166. The images of these
  surveys are based on photographic data obtained using the Oschin
  Schmidt Telescope on Palomar Mountain and the UK Schmidt Telescope.
  The plates were processed into the present compressed digital form
  with the permission of these institutions.

  This research has made use of the NASA/IPAC Extragalactic Database
  (NED) which is operated by the Jet Propulsion Laboratory, California
  Institute of Technology, under contract with the National
  Aeronautics and Space Administration.

\end{acknowledgements}


\end{document}